\documentclass[12pt]{article}
\usepackage{amsmath,amssymb,color}

\newcommand{\bea}{\begin{eqnarray}}
\newcommand{\eea}{\end{eqnarray}}
\newcommand{\be}{\begin{equation}}
\newcommand{\ee}{\end{equation}}
\newcommand{\vs}[1]{\vspace{#1 mm}}

\newcommand{\dsl}{\pa \kern-0.5em /}

\newcommand{\pa}{\partial}

\newcommand{\nn}{\nonumber\\}

\begin{document}
\topmargin 0mm
\oddsidemargin 0mm

\begin{flushright}

USTC-ICTS/PCFT-20-05\\

\end{flushright}

\vspace{2mm}

\begin{center}

{\Large \bf More on the open string pair production }

\vs{10}

{\large  J. X. Lu and Nan Zhang}

\vspace{4mm}

{\em  Interdisciplinary Center for Theoretical Study\\
 University of Science and Technology of China, Hefei, Anhui
 230026, China\\
 \medskip
 Peng Huanwu Center for Fundamental Theory, Hefei, Anhui 230026, China\\

}

\end{center}

\vs{10}

\begin{abstract}
 Motivated by the recent work \cite{Lu:2019ynq} by one of the present authors, we here report that there exist two new systems, namely, D3/(F, D1) and D3/(D3, (F, D1)), either of which can give rise to a much larger sizable open string pair production rate at  a  condition much relaxed.  Here the D3 is taken as our own (1 + 3)-dimensional world  while the non-threshold bound state  (F, D1) or (D3, (F, D1)) is placed parallel nearby in the directions transverse to both our D3 world and the non-threshold bound state considered.
 \\
\end{abstract}

\newpage

\section{Introduction}

In a series of publications by one of the present authors and his collaborators \cite{Lu:2017tnm, Lu:2018suj, Lu:2018nsc, Jia:2018mlr},   it was uncovered  the existence of  open string pair production, in the spirit of Schwinger pair production in QED \cite{Schwinger:1951nm},   for a system of two Dp branes in Type II superstring theories\footnote{An isolated Dp brane, even it carries a constant electric flux, cannot give rise to the open string pair production, unless the applied electric flux reaches its critical value and can break the open strings.  The reason for this is that the strings in Type II string theory are oriented and as such an open string in this theory is charge neutral.  In other words, the two ends of an open string must carry an equal but opposite sign of charge and its net charge is zero.  However, this is not the case for un-oriented  bosonic string or un-oriented Type I superstring and the corresponding open string pair production was discussed a while ago by Bachas and Porrati in \cite{Bachas:1992bh, Porrati:1993qd}.}.   Here the two Dp-branes  are placed parallel at a separation with each carrying a constant worldvolume electric flux. In particular, the pair production rate can be greatly enhanced in the presence of certain constant magnetic flux in addition to the applied electric flux and the largest possible rate occurs for $p = 3$ given the smallness of the earthbound realizable electric and magnetic fluxes with respect to string scale \cite{Lu:2017tnm,  Lu:2018nsc, Jia:2018mlr}.  Even so, the rate itself is still too small for any practical use.  To have a large rate, one choice is to have a large magnetic flux which needs to be on the order of string scale.   This string scale magnetic flux certainly cannot be realized in an earthbound laboratory.   In a recent publication by the same author \cite{Lu:2019ynq}, we report a sizable open string pair production rate when the brane separation reaches a few times of string scale for a system of D3 brane, taken as our own (1 + 3)-dimensional world and carrying an electric flux, and a nearby D1 brane, placed parallel in directions transverse to both branes.  For this system,  the D1 effectively plays  the role of a string scale magnetic flux with respect to the D3 \cite{Lu:2009pe, Breckenridge:1996tt, Costa:1996zd, Di Vecchia:1997pr}. So a large rate can indeed be realized for such a system but the unwanted feature is that this occurs almost at the brane separation for the onset of tachyon condensation.  Note that the brane interaction is also attractive at least for fairly large brane separation.

In spite of the above, the large transient pair production and the following tachyon condensation along also with the pair production will release excess energy of the underlying system before it settles down to a stable configuration in a very short period of time (expected also in the string scale)  in terms of, for example, radiation.  If string theory is indeed relevant to our real world, this short time energy burst may give rise to the so-called $\gamma$-ray burst if it occurs in the late time of our universe or may provide an alternative mechanism for reheating process if it happens, for example, during the early stage of our universe.  Exploring either of these two possibilities is worthwhile and this may be our future plan. In this paper, our focus is however limited to find new systems which can provide an improvement of the aforementioned condition used for the D3/D1 for a sizable pair production rate and at the same time can give rise to an even larger sizable pair production rate.  The D1 in the D3/D1 plays already the effective string scale magnetic flux to our own D3 as mentioned above.  In order to give a large pair production rate, we need also a string scale electric flux. Surely we cannot do much on our own D3. We know that in non-perturbative string theories,  in addition to the 1/2 BPS simple D-branes, we have also the so-called 1/2 BPS non-threshold bound states,  some of which can carry a constant string scale electric flux.  So for this purpose, we can replace the D1 in D3/D1 by such a 1/2 BPS non-threshold bound state. It turns out, in this respect, that either the 1/2 BPS (F, D1) non-threshold bound state or the 1/2 BPS (D3, (F, D1)) non-threshold bound state does the work.  As will be seen, this will provide solution to both of the issues mentioned above, namely relaxing the condition for a sizable rate and giving a much larger sizable rate at the same time.

In other words, we will replace the D1 in the system of D3/D1 mentioned above by the (F, D1) or (D3, (F, D1)) in this paper.  Here (F, D1) stands for the non-threshold bound state of fundamental strings and D1-branes \cite{Schwarz:1995dk,Witten:1995im} with their respective co-prime quantized integral charge ($p', q'$). From the viewpoint of the D1 world-sheet, the F-strings are given by a quantized constant electric flux \cite{Witten:1995im, Callan:1997kz, DiVecchia:1999uf,Lu:2009yx}. (D3, (F, D1)) is the non-threshold bound state of D3-branes and the delocalized non-threshold bound state (F, D1) \cite{Lu:1999uv, Lu:2000ys}, characterized by three integers ($n', (p', q')$) without a common divisor.  From the D3 brane worldvolume viewpoint, the delocalized (F, D1) non-threshold bound states are given by collinear quantized constant electric and magnetic fluxes.

 This paper is organized as follows. In  the following  section, we will focus on the system of D3/(F, D1).  In section 3, we will move to discuss the system of D3/(D3, (F, D1)). We will discuss and conclude this paper in section 4.

 \section{The open string pair production: D3/(F, D1)}

For this system, we consider that for general purpose our D3 carries earthbound realizable collinear electric flux $\hat f$ and magnetic flux $\hat g$, both of which are extremely small, while  the non-threshold bound state (F, D1) can be viewed as  D1 carrying a quantized electric flux $\hat f'$ as follows\footnote{We here choose our own D3 along $x^{1}, x^{2}, x^{3}$-directions while the D1 (or (F, D1)) is along $x^{1}$-direction with a transverse separation from the D3.}

\be\label{flux}
\hat F_{3} = \left(\begin{array}{cccc}
0& -\hat f&0&0\\
\hat f&0&0&0\\
0&0&0&- \hat g\\
0&0& \hat g&0\end{array}\right), \qquad \hat F'_{1} = \left(\begin{array}{cc}
0 & \hat f'\\
- \hat f'& 0\end{array}\right),
\ee
where both $\hat F_{3}$ and $\hat F'_{1}$ are dimensionless and the quantized electric flux
\be\label{quantizedef}
\hat f' = \frac{p'}{\sqrt{p'^{2} + q'^{2}/g^{2}_{s}}},
\ee
with  $p'$ and $q'$ two co-prime integers.  Here $q'$ denotes the multiplicity of D1-branes and $p'$ the number of F-strings, and $g_{s}$ the string coupling. We denote also $\Delta_{(p', q')} = p'^{2} + q'^{2}/g^{2}_{s}$ for simplicity. Note that we need to keep $g_{s} \ll 1$ such that the D-branes used can be treated as rigid ones at least to the probe distance not much smaller than the string scale $l_{s}$ as discussed in \cite{Jia:2019hbr}. In the remaining of this paper, we take $g_{s} = 10^{- 2}$ as a sample small coupling for later concrete discussion.

To compute the pair production rate, we first compute the closed string tree-level cylinder amplitude between the D3 and the D1.  For this, a trick was employed in \cite{Jia:2019hbr} via an extension of the D1 to a D3 carrying the following flux,
\be
\hat F'_{3} = \left(\begin{array}{cccc}
0 & \hat f'&0&0\\
- \hat f'& 0&0&0\\
0&0&0&- \hat g'\\
0&0& \hat g'&0
\end{array}\right),
\ee
where the magnetic flux $g'$ is taken to be infinity at the end of computations.  The basic picture behind is that this magnetic flux  gives rise to the delocalized D1 within the new D3.  We actually have a non-threshold bound state (D3, (F, D1)).  Our own D3, when carrying also electric and magnetic fluxes as given in (\ref{flux}), is a non-threshold bound state (D3, (F, D1)), too.  So the interaction now is in fact between the two non-threshold bound states which is actually the focus of the following section.  However, when we take $\hat g' \to \infty$, this interaction is dominated by that between our own (D3, (F, D1)) and the delocalized (F, D1) of the extended (D3, (F, D1)) and can be reduced by a procedure as discussed in detail in \cite{Jia:2019hbr} to that between our own (D3, (F, D1)) and the (F, D1).

This closed string cylinder amplitude can then be read from the last equality of the general one given in  (136) in \cite{Jia:2019hbr}  for $p = 3, p' = 1$\footnote{Note that here $p'$ stands for the spatial dimensions of Dp$'$ brane, not the previously mentioned quantized integral charge of F-strings.}  with $\nu_{2} = 0$ and the respective fluxes given in (\ref{flux}) as
\bea\label{c-ampli}
&&\Gamma_{3, 1} = \frac{2 q'\, V_{2} \sqrt{(1 - \hat f'^{2})(1 - \hat f^{2})(1 + \hat g^{2})} (\cosh \pi \bar\nu_{0} - \cos\pi \nu_{1})^{2}}{8 \pi^{2} \alpha'} \int_{0}^{\infty} \frac{d t}{t^{3}} ~e^{- \frac{y^{2}}{2 \pi \alpha' t}} \nn
&&\times \prod_{n = 1}^{\infty} \frac{[1 - 2 |z|^{2n} e^{-\pi \bar\nu_{0}} \cos\pi\nu_{1} + e^{- 2\pi \bar\nu_{0}} |z|^{4n}]^{2} [1 - 2 |z|^{2n} e^{\pi \bar\nu_{0}} \cos\pi\nu_{1} + e^{ 2\pi \bar\nu_{0}} |z|^{4n}]^{2} }{(1 - |z|^{2n})^{4} [1 - 2 |z|^{2n} \cosh2\pi \bar\nu_{0}
+ |z|^{4n}] [1 - 2 |z|^{2n} \cos2\pi\nu_{1}
+ |z|^{4n}]},
\eea
where $y$ is the brane separation along the directions transverse to both D3 and D1, $|z| = e^{- \pi t} < 1$, $V_{2}$ the volume of D1 worldvolume, $\alpha'$ the Regge slope parameter, and the so-called electric parameter $\bar\nu_{0}$ and the magnetic parameter $\nu_{1}$ are determined by the electric fluxes ($\hat f, \hat f'$) and the magnetic one ($\hat g$), respectively, via
\be\label{nu-parameter}
\tanh \pi \bar\nu_{0} = \frac{|\hat f - \hat f'|}{1 - \hat f \hat f'}, \qquad \tan \pi \nu_{1} = \frac{1}{\hat g},
\ee
with $\bar\nu_{0} \in [0, \infty), \nu_{1} \in (0, 1)$.  In the amplitude (\ref{c-ampli}), we have added an integer factor $q'$, counting the multiplicity of D1-branes.

Note that the dimensionless flux (denoted with a hat above) here is defined via $\hat F = 2\pi \alpha' F$ with $F$ the usual dimensionful field strength (without a hat above).  So we have in general $|\hat g| \in [0, \infty)$ and $|\hat f|, |\hat f'| \in [0, 1)$ with unity here denoting the  critical value of the respective electric flux. Unlike what we did in \cite{Lu:2019ynq} where $\hat f' = 0$ is chosen, we here consider $\hat f' = p'/\Delta^{1/2}_{(p', q')}$ which can  be much larger than the small earthbound  $\hat f$ even for small integers $p'$ and $q'$ as well as a small string coupling $g_{s}$. We will illustrate this numerically later on.  In other words,  the electric parameter $\bar\nu_{0}$, from (\ref{nu-parameter}), can be completely determined by this flux in practice and is much larger than the one considered in \cite{Lu:2019ynq}.   This certainly enhances the open string pair production rate.

Note also that  the above amplitude (\ref{c-ampli}) is strictly positive for large $y$ and this remains true even if we turn off our D3 worldvolume fluxes,  i.e., by setting $\hat f = 0, \hat g = 0$ (therefore $\bar \nu_{0} \neq  0, \nu_{1} = 1/2$ from (\ref{nu-parameter})), hence giving an attractive interaction between the D3 and the (F, D1) according to our conventions\footnote{The positive amplitude gives an attractive interaction while a negative one gives a repulsive interaction.}. This indicates that the underlying system does not preserve any supersymmetry, consistent with the known fact. We can understand the amplitude being positive for large $y$ the following way.  Note that the dominant contribution to the amplitude for large $y$ comes from the large $t$ integration, due to the exponentially suppressing factor ${\rm Exp} [- y^{2} /(2\pi\alpha' t)]$ in the integrand, and as such  every factor in the integrand is positive, so giving a positive amplitude.  However, the nature of the amplitude is obscure for small $y$. Now the small $t$ integration becomes also important and the factor $[1 - 2 |z|^{2n} \cosh 2\pi \bar \nu_{0} + |z|^{4n}] \approx 2 (1 - \cosh 2\pi\bar\nu_{0}) $  in the denominator of the infinite product in the integrand of (\ref{c-ampli}) can be negative for small $t$.  Once this happens, the sign of the amplitude is ambiguous since there are an infinite number of such factors in the product.  This ambiguity actually indicates a potential new physics to occur and the best way to decipher this is to pass the closed string cylinder amplitude to the corresponding open string one-loop annulus one via a Jacobi transformation\footnote{Certain relations for the Dedekind $\eta$-function and the $\theta_{1}$-function have also been used, see \cite{Jia:2019hbr} for detail, for example.} by sending $t \to 1/t$. The resulting annulus amplitude is
\bea\label{annulus-amplit}
\Gamma_{3, 1} &=& \frac{2 \,q'\, V_{2} |\hat f - \hat f'| }{8 \pi^{2} \alpha'} \int_{0}^{\infty} \frac{d t}{t} ~e^{- \frac{y^{2} t}{2 \pi \alpha' }} \frac{[\cosh \pi\nu_{1} t - \cos\pi\bar\nu_{0} t]^{2}}{\sin\pi\bar\nu_{0} t \sinh\pi\nu_{1} t}\qquad\nn
&\,& \times \prod_{n = 1}^{\infty} \frac{|1 - 2 |z|^{2n} e^{- i \pi \bar\nu_{0} t} \cosh \pi \nu_{1} t  + |z|^{4n} e^{- 2 i \pi \bar\nu_{0} t}|^{4}}{(1 - |z|^{2n})^{4} [1 - 2 |z|^{2n} \cosh 2\pi \nu_{1} t + |z|^{4n}][1 - 2 |z|^{2n} \cos2\pi\bar\nu_{0} t + |z|^{4n}]},
\eea
where $|z| = e^{- \pi t}$ continues to hold.  Except for the factor $\sin\pi\bar\nu_{0} t$, all other factors in the integrand are positive since $1 - 2 |z|^{2n} \cos 2\pi\bar\nu_{0} t + |z|^{4n} > 1 - 2 |z|^{2n} + |z|^{4n} = (1 - |z|^{2n})^{2} > 0$ and
$1 - 2 |z|^{2n} \cosh2\pi \nu_{1} t + |z|^{4n} = (1 - e^{2\pi \nu_{1} t} |z|^{2n}) (1 - e^{- 2\pi \nu_{1} t} |z|^{2n}) > 0$ since $\nu_{1} < 1$ and $n \ge 1$.  In particular, the integrand gives a potential tachyonic instability when $t \to \infty$ (corresponding to the closed string $t \to 0$ since the two are inversely related to each other) since it blows up
\be\label{tachyon}
\sim e^{- \frac{y^{2} t}{2 \pi \alpha' }} e^{\pi \nu_{1} t} = e^{- 2\pi \alpha' t \left[\frac {y^{2}} {(2\pi \alpha')^{2}} - \frac{ \nu_{1}} {2 \alpha'} \right]},
\ee
when $y < \pi \sqrt{2 \nu_{1} \alpha'}$. This is due to the existence of the so-called tachyonic shift $\nu_{1}/2$ which will be discussed further later on.  As we will see, this shift will also give rise to the enhancement of the open string pair production. The aforementioned factor $\sin\pi\bar\nu_{0} t$ actually gives rise to an infinite number of simple poles of the integrand at $t_{k} = k/\bar\nu_{0}$ with $k = 1, 2, \cdots$ along the positive t-axis, reflecting the existence of an imaginary part of the amplitude.  This imaginary part indicates the decay of the underlying system via the so-called open string pair production.  The decay rate can be computed as the sum of the residues of the integrand in (\ref{annulus-amplit}) at these poles times $\pi$ per unit worldvolume following  \cite{Bachas:1992bh} as
\be\label{decay-rate}
{\cal W} = \frac{4\, q'\, |\hat f - \hat f'|}{8 \pi^{2} \alpha'}\sum_{k = 1}^{\infty} \frac{(- )^{k - 1}}{k} e^{- \frac{k y^{2}}{2\pi \alpha' \bar\nu_{0}}} \frac{\left[\cosh \frac{\pi k \nu_{1} }{\bar\nu_{0}} - (-)^{k}\right]^{2}}{\sinh \frac{\pi k \nu_{1} }{\bar\nu_{0}}} Z_{k} (\bar\nu_{0}, \nu_{1}),
 \ee
where \be\label{Zk}
Z_{k} (\bar\nu_{0}, \nu_{1}) = \prod_{n =1}^{\infty} \frac{\left(1 - 2 (-)^{k} |z_{k}|^{2n} \cosh \frac{\pi k \nu_{1} }{\bar\nu_{0}} + |z_{k}|^{4n} \right)^{4}}{(1 - |z_{k}|^{2n})^{6} \left(1 - 2 |z_{k}|^{2n} \cosh \frac{2\pi k \nu_{1} }{\bar\nu_{0}} + |z_{k}|^{4n}\right)},
\ee
with $|z_{k}| = e^{- \pi k/\bar\nu_{0}}$.
Following \cite{nikishov},  the rate for the open string pair production corresponds just to the leading $k = 1$ term of the above decay rate and it is
\be\label{pprate}
{\cal W}^{(1)} = \frac{4\, q'\, |\hat f - \hat f'|}{8 \pi^{2} \alpha'} e^{- \frac{y^{2}}{2\pi \alpha' \bar\nu_{0}}} \frac{\left[\cosh \frac{\pi  \nu_{1} }{\bar\nu_{0}} + 1\right]^{2}}{\sinh \frac{\pi  \nu_{1} }{\bar\nu_{0}}} Z_{1} (\bar\nu_{0}, \nu_{1}),
\ee
where
\be\label{Z1}
Z_{1} (\bar\nu_{0}, \nu_{1}) = \prod_{n =1}^{\infty} \frac{\left[1 + 2  |z_{1}|^{2n} \cosh \frac{\pi  \nu_{1} }{\bar\nu_{0}} + |z_{1}|^{4n} \right]^{4}}{(1 - |z_{1}|^{2n})^{6} \left[1 - 2 |z_{1}|^{2n} \cosh \frac{2\pi  \nu_{1} }{\bar\nu_{0}} + |z_{1}|^{4n}\right]}.
\ee
With the above computed open string pair production rate, we can now discuss how large this rate can be. Suppose that the D3 is the (1 + 3)-dimensional world we are living in and there exists a nearby (F, D1) in the directions transverse to our D3.
For practical purpose, we can only control the sizes of the collinear electric and magnetic fields on our own D3, both of which are in general  very small compared with the string scale. The possible largest static electric field\footnote{\label{fnnew}  We  consulted our experimental colleague Zhengguo Zhao and learned that the current laboratory limit for electric field is on the order of $10^{10} \, {\rm Volt}/m$. The strongest direct-current magnetic field generated is on the order  of  $50$ Tesla, see \cite{smf} for example.  This gives $e E\sim e B \sim 10^{- 8} m^{2}_{e}$ with $m_{e}$ the electron mass. This electric field is still eight orders of magnitude smaller than that required for giving rise to the Schwinger pair production in QED.} which can be realized in an earthbound laboratory is $E \sim 10^{10}\, {\rm Volt/m}$ which gives $e E \sim 10^{- 8} \,m^{2}_{e} \sim 2.5 \times 10^{- 21}\, {\rm TeV}^{2}$ with $m_{e}$ the electron mass (The largest static magnetic field gives also $e B \sim 10^{- 8}\, m^{2}_{e} \sim 2.5 \times 10^{-21} \, {\rm TeV}^{2}$, see footnote (\ref{fnnew})).  Note that with the experimental constraints the bound for the string scale $M_{s} =  1/l_{s}$ can be a few TeV upto $10^{16} \sim 10^{17}$ GeV , see \cite{Berenstein:2014wva}, for example.  So even the lowest bound $\sim$ a few TeV for the string scale is taken, we have $\hat f = 2\pi \alpha' e E \sim 2 \pi e E/M^{2}_{s} \le 10^{- 21} \ll 1$ and $ \hat g = 2\pi \alpha' e B \sim 10^{- 21} \ll 1$. So indeed we have both $\hat f \ll 1$ and $\hat g \ll 1$ in practice.

Given  $\hat g \le 10^{- 21} \ll 1$,  we have  $\nu_{1} \lesssim 1/2$ from (\ref{nu-parameter}).  Note that the quantized electric flux $\hat f' < 1$ from (\ref{quantizedef}) since $p', q' \ge 1$ but it is still in general much larger than the applied $\hat f \le  10^{- 21}$. So we can  completely ignore $\hat f$ in general and have from (\ref{quantizedef})
\be\label{effective-nu0}
\tanh\pi \bar\nu_{0} \approx  \hat f' = \frac{g_{s} p'}{\sqrt{q'^{2} + g_{s}^{2} p'^{2}}}.
\ee
In general, the larger the electric flux $\hat f'$ is, the larger the pair production rate. This can be easily seen from the rate given in (\ref{pprate}) since the exponential factor ${\rm Exp}[- \frac{y^{2}}{2\pi \alpha' \bar\nu_{0}}]$ dominates the rate among other things. It is clear from (\ref{effective-nu0}) that the largest possible $\hat f'$ can be reached for $q' = 1$ (Note that $q' \ge 1$). For not too large $p'$ (see \cite{Copeland:2003bj}), say  $p' = 10$ for example, we have $\hat f' \approx g_{s} p' = 0.1$ if $g_{s} = 10^{-2}$ as given earlier.  We can also write $\hat f' = 2 \pi \alpha'  e E' = g_{s} p'$, giving $e E' = g_{s} p' /(2\pi \alpha') = M_{s}^{2}/(20 \pi)$. We have then, from (\ref{effective-nu0}), $\pi \bar \nu_{0} \approx g_{s} p' \to \bar\nu_{0} = g_{s} p' /\pi = 1/(10 \pi)$. With this $\bar \nu_{0}$,  we have $|z_{1}| = e^{- \pi/\bar\nu_{0}} = e^{- 10 \pi^{2} } \ll 1$,  which implies the factor (\ref{Z1}) $Z_{1} (\bar\nu_{0}, \nu_{1})  \approx 1$ for the rate (\ref{pprate}).  In other words, only the lowest-mass modes of the open string connecting the D3 and the D1, with each having mass $m = y/(2\pi \alpha')$, contribute to the rate (\ref{pprate}) when $|z_{1}| \ll 1$.  The rate (\ref{pprate}) becomes now
\be\label{eff-pprate}
{\cal W}^{(1)} \approx  \frac{e E'}{2 \pi} e^{- \pi \frac{m^{2}_{\rm eff}}{e E'}},
\ee
where $e E' = M^{2}_{s}/ (20 \pi)$ and we have defined an effective mass
\be \label{effm}
m^{2}_{\rm eff} \equiv m^{2} - \frac{\nu_{1}}{2\alpha'},
\ee
with $m = y/(2\pi \alpha')$ and $\nu_{1} \lesssim 1/2$ as given in (\ref{nu-parameter}).  Note that this rate is computed as the number of pairs produced per unit time and per unit D-string length.

Before we come to discuss this rate in detail, let us first understand which modes actually contribute to this rate. For this and the tachyonic shift mentioned earlier,   we extend the D1 to the D3 as prescribed earlier with a large $\hat g'$ (we will take it to be infinity at the end).   We then have
\be\label{nu1}
\tan \pi \nu_{1} = \frac{|\hat g - \hat g'|}{1 + \hat g \hat g'},
\ee
which gives its correspondence in (\ref{nu-parameter}) when we take $\hat g' \to \infty$. In the absence of worldvolume fluxes,  the mass spectrum of the open superstring connecting the two D3 is given as
 \be\label{mass-level}
\alpha' M^{2} = -\alpha' p^{2} =  \left\{\begin{array}{cc}
\frac{y^{2}}{4\pi^{2} \alpha'} + N_{\rm R} &  (\rm R-sector),\\
\frac{y^{2}}{4\pi^{2} \alpha'}  + N_{\rm NS}  - \frac{1}{2} & (\rm NS-sector),
\end{array}\right.
\ee
where $p = (k, 0)$ with $k$ the momentum along the brane worldvolume directions, $N_{\rm R}$ and $N_{\rm NS}$ are the standard number operators in the R-sector and NS-sector, respectively, as
\bea\label{NP}
N_{\rm R} &=& \sum_{n = 1}^{\infty} (\alpha_{- n} \cdot \alpha_{n} + n d_{-n} \cdot d_{n}),\nn
N_{\rm NS} &=& \sum_{n = 1}^{\infty} \alpha_{- n} \cdot \alpha_{n} + \sum_{r = 1/2}^{\infty} r d_{- r} \cdot d_{r}.
\eea
When $y = 0$, the supersymmetric open string/anti open string each has 16 massless modes, which are eight bosons ($8_{\rm B}$) and eight fermions ($8_{\rm F}$), in addition to an infinite tower of massive modes, from the 10 dimensional viewpoint.  From the D3 brane viewpoint,  these massless modes when $y = 0$ just give rise to the unbroken N = 4 supersymmetric and broken U(1)  massive super Yang-Mills modes when $y \neq 0$, which are five massive scalars, four massive spinors and one massive vector in 4-dimensions. As discussed in detail in \cite{Lu:2019ynq}, the original U(2) when $y = 0$ is broken to U(1) $\times$ U(1) when $y \neq 0$ and for the two broken generators, one gives the mass $m = y/(2\pi \alpha')$ for the modes associated with the open string and the other gives the same mass for the modes associated with the anti open string.  Each of these modes from the open string carries, say, a positive unity charge while each of the modes from the anti open string carries a negative unity charge under the unbroken U(1) associated with our own D3. So totally we have 16 pairs of charged/anti-charged modes.  Among these, we have five  pairs of charged/anti-charged massive scalars, four pairs of charged/anti-charged massive spinors (counting 8 pairs of charged/anti-charged polarizations) and one pair of charged/anti-charged vectors (counting 3 pairs of charged/anti-charged polarizations).  We try to see which pair or pairs of charged/anti charged polarizations to contribute to the pair production rate (\ref{eff-pprate}).

For this, we now turn on  a small magnetic flux $\hat g \ll 1$ on our own D3 and a large $\hat g'$ on the other D3 which is taken effectively as the D1 when $\hat g' \to \infty$ mentioned earlier.  The energy in the R-sector is strictly positive and the R-sector vacuum, consisting of massive $8_{\rm F}$ or 4 massive four-dimensional Majorana spinors when $y \neq 0$, has no the so-called tachyonic shift, just like the QED case \cite{Ferrara:1993sq,Bolognesi:2012gr}.
The NS-sector has a different story, however. The energy spectrum in the NS-sector, following \cite{Ferrara:1993sq,Bolognesi:2012gr}, is now\footnote{This energy spectrum is for open string carrying positive unit charge at its one end of interest and the spin polarization is along the direction of magnetic field.  For anti open string carrying negative unity charge, the same spectrum holds for its spin polarization along the opposite direction.}
\be\label{OSM}
\alpha' E^{2}_{\rm NS} =  (2 N + 1) \frac{\nu_{1}}{2} - \nu_{1} \, S + \alpha' M^{2}_{\rm NS},
\ee
where the magnetic parameter $\nu_{1}$ is defined in (\ref{nu1}),  the Landau level  $N = b^{+}_{0} b_{0} $, the spin operator in the 23-direction is
\be\label{spin}
S = \sum_{n = 1}^{\infty} (a^{+}_{n} a_{n} - b^{+}_{n} b_{n}) + \sum_{r = 1/2}^{\infty} (d^{+}_{r} d_{r} - \tilde d^{+}_{r} \tilde d_{r}),
\ee
and the mass $M_{\rm NS}$ in the NS-sector is
\be\label{new-mass}
\alpha' M^{2}_{\rm NS}  = \frac{y^{2}}{4 \pi^{2} \alpha'} + N_{\rm NS} - \frac{1}{2},
\ee
but now with
\be
N_{\rm NS} = \sum_{n = 1}^{\infty}  n (a^{+}_{n} a_{n} + b^{+}_{n} b_{n}) + \sum_{r = 1/2}^{\infty} r (d^{+}_{r} d_{r} + \tilde d^{+}_{r} \tilde d_{r})
+ N^{\perp}_{\rm NS},
\ee
where $N^{\perp}_{\rm NS}$ denotes the contribution from directions other than the 2 and 3 ones.  The only states which have the potential to be tachyonic \cite{Ferrara:1993sq}, in particular for $y = 0$,  belong to the first Regge trajectory, given by
\be\label{fRt}
(a^{+}_{1})^{\tilde n} d^{+}_{1/2} |0 \rangle_{\rm NS},
\ee
where  $|0 \rangle_{\rm NS}$ is the NS-sector vacuum,  projected out in the superstring case. For these states, we have
\be\label{EL}
\alpha' E^{2}_{\rm NS} = - \frac{\nu_{1}}{2} + (1 - \nu_{1}) (S - 1) +  \frac{y^{2}}{4 \pi^{2} \alpha'},
\ee
where $S = \tilde n + 1 \ge 1$.  In the absence of magnetic fluxes, the above gives the mass in the NS-sector as
\be\label{ML}
\alpha' M^{2}_{\rm NS} = \frac{y^{2}}{4 \pi^{2} \alpha'} + S - 1,
\ee
which  becomes massless  for the spin $S = 1$ state $d^{+}_{1/2} |0\rangle_{\rm NS}$ when $y = 0$  and is massive for $y \neq 0$.    A finite brane separation $y$ for the spin $S = 1$ state $d^{+}_{1/2} |0\rangle_{\rm NS}$ corresponds to the gauge symmetry breaking $U(2)  \to U(1) \times U(1)$, with the W-boson mass $M_{W} = y/(2\pi \alpha')$,   as mentioned earlier.  From (\ref{EL}) and (\ref{ML}), the lowest mass is for the spin $S = 1$ state and the state $d^{+}_{1/2} |0\rangle_{\rm NS}$ has its energy
\be\label{energy}
\alpha' E^{2}_{\rm NS} = - \frac{\nu_{1}}{2} +  \frac{y^{2}}{4 \pi^{2} \alpha'}.
\ee
 This energy is precisely the effective mass defined for the rate (\ref{eff-pprate}) given in (\ref{effm}). We also see the  tachyonic shift mentioned right after (\ref{tachyon})  in the presence of magnetic flux\footnote{Adding an additional magnetic flux, say, along 45-direction with the corresponding magnetic parameter $\nu_{2} > 0$, without loss of generality assuming $\nu_{2} \le \nu_{1}$, will actually reduce rather than increase the shift to $(\nu_{1} - \nu_{2})/2$. So this will diminish rather than increase the rate \cite{Jia:2018mlr}.}. It is clear that a tachyonic instability can occur when the brane separation $y \le \pi \sqrt{2 \nu_{1} \alpha'}$.  Once this happens, we will have a phase transition via the so-called tachyon condensation \cite{Ferrara:1993sq}. From the viewpoint of the D3 worldvolume, this instability is the Nielsen-Olesen one for the non-abelian gauge theory of the 4-dimensional N = 4  U(2) super Yang-Mills\cite{Nielsen:1978rm} in the so-called weak field limit.

The above discussion clearly indicates that only the pair of charged/anti charged massive vector polarizations specified above, among the 16 pairs of charged/anti charged polarizations mentioned earlier, contributes to the pair production rate (\ref{eff-pprate}).  This implies that the contribution of all the other pairs to the rate is vanishingly small with respect to this one.  The precise reason for this is the intrinsic $\nu_{1} = 1/2$ associated with the D1\footnote{It is well-known that an open string with its two ends carrying charges placed in  a magnetic flux will have its mass splittings due to the Landau motion in this magnetic background and the spin coupling with the magnetic flux,  just like a charged particle with spin moving in a given magnetic field.   In particular, these massive charged modes, with respective to the D3 brane observer, appear as massive charged particles.   Because of these splittings, different mode can have its different contribution to the pair production rate.} which can be taken effectively as the $\hat g' \to \infty$ limit of the other D3 in the absence of magnetic flux on our own D3. For this pair of polarizations, its contribution to the rate has a factor $e^{\pi \nu_{1}/(2 \alpha' e E')} \approx e^{5 \pi^{2}} \gg 1$. The contribution of any other pair of polarizations to the rate is at least smaller by this factor.   From the 4-dimensional perspective, this pair of vector polarizations consists of the positive unit charge vector polarization along the magnetic flux direction and the negative unit charge vector polarization opposite to the magnetic flux direction.

Note that the system D3/(F, D1) breaks all the spacetime supersymmetries. This is consistent with the energy splittings of the charged polarizations discussed above due to
the $\nu_{1} = 1/2$ (Otherwise, we expect all the 16 modes from the open string along with the other 16 modes from the anti open string all have the same mass given as $y/(2\pi \alpha')$).

We now come to discuss the other properties of the rate (\ref{eff-pprate}).  This rate formula looks almost the same as that given in \cite{Lu:2019ynq} for the D3/D1 system except for one important difference. For example,  the effective mass in both cases is essentially the same as that\footnote{Since the laboratory magnetic flux $\hat g \sim 10^{-21} \ll 1$, we always have $\nu_{1} \sim 1/2$ from (\ref{nu-parameter}), essentially independent of the laboratory magnetic flux $\hat g$.} given in (\ref{effm}) with $\nu_{1} \approx 1/2$. However, their sharp difference is the electric field appearing in the rate formula (\ref{eff-pprate}). In the present case, the electric field $E'$ is the one given by the quantized electric flux $\hat f' = 2\pi \alpha' e E' = g_{s} p$, giving $e E' = M^{2}_{s}/(20 \pi)$, while in  \cite{Lu:2019ynq}, the electric field is the laboratory one with  its current largest possible value $e E \sim 10^{- 8}\, m^{2}_{e} \sim 10^{-21} \, {\rm TeV}^{2}$ which is far more smaller than the former.  In other words, we have $e E' \gg e E$. Precisely because of this, it is much easier to produce the pair production from the present D3/(F, D1) system than from the D3/D1.

In order to produce the charged pairs in reality, we need an electric field which is large enough to separate the virtual open string and the anti open string (or the charged/anti charged pair viewed from the brane world), once they are created, over a distance of order of their Compton length.  We can estimate this by setting the work done by the electric force acting on the charged/anti charged pair to separate them over their Compton length $1/m_{\rm eff}$, i.e $2 e E' /m_{\rm eff}$, to equal their rest energy  $2\, m_{\rm eff}$. This gives the standard condition to actually produce the pair as
\be\label{detectc}
e E' \approx m^{2}_{\rm eff}.
\ee
Note  that $e E' = M^{2}_{s}/(20 \pi)$, not much smaller than the string scale, and $m^{2}_{\rm eff}$ from (\ref{effm}) can range from small to the string scale through changing the brane separation so long $m^{2}_{\rm eff} > 0$ is kept. So the above condition can be easily satisfied for the present case.  In the estimation, we can actually take
$\nu_{1} \approx 1/2$ since $\hat f \ll 1$ and $\hat g  \ll 1$ and both can be ignored in practice.  From the condition $m^{2}_{\rm eff} > 0$ to avoid tachyon condensation and $e E' \ge m^{2}_{\rm eff}$ to guarantee the pair production,  we can estimate the range of brane separation to be $y_{0} \le y \le y_{0} + 0.1\, l_{s}$ with $y_{0} = \pi l_{s}$, independent of the string scale in the following sense. However, the similar range of the brane separation for  the system of D3/D1 reported in \cite{Lu:2019ynq} is $y_{0} \le y \le y_{0} + 2\pi \times 10^{-8} \, (m_{e}/M_{s})^{2} l_{s}$ which depends on the string scale.  For example, even if we take the lower bound for the string scale $M_{s} =$ a few TeV \cite{Berenstein:2014wva},  we will have the range as $y_{0} < y  < y_{0} + 10^{-21} l_{s}$.  It  is so small and makes the pair production hardy possible before the tachyon condensation to occur.  For the present D3/(F, D1) system, the range is $0.1 l_{s}$ and is independent of the string scale. Though this range is still small,  it is much larger than the one for the D3/D1 system and so has a great chance to produce the pair before the tachyon condensation.  Given what has been said,  the detection of this pair production is most likely remote since the time duration for this to occur before the tachyon condensation is also expected to be on the order of one tenth of string scale.  This transient process is expected since the range of brane separation is $0.1 l_{s}$ and the brane interaction is attractive.

We now move to the other new system which has also advantage features in terms of the pair production over the D3/D1 one.

\section{The open string pair production: D3/(D3, (F, D1))}

We now move to discuss the other system  D3/(D3, (F, D1)) which can also provide a better choice for producing the open string pairs than the D3/D1 one in a similar spirit as discussed in the previous section.  Here again the D3 is assumed to be our (1 + 3)-dimensional world. The non-threshold bound state (D3, (F, D1)) \cite{Lu:1999uv} is placed parallel to our D3 braneworld at a separation $y$ along the directions transverse to both.  Our D3 carries the same worldvolume flux $\hat F_{3}$ as given in (\ref{flux}) in the previous section.  The non-threshold bound state (D3, (F, D1)) \cite{Lu:1999uv, Lu:2000ys}  can be viewed as $n'$ D3 branes carrying their following worldvolume quantized electric and magnetic flux $\hat F'_{3}$

\be\label{flux'}
\hat F'_{3} = \left(\begin{array}{cccc}
0& \hat f'&0&0\\
-\hat f'&0&0&0\\
0&0&0&\hat g'\\
0&0&- \hat g'&0\end{array}\right),
\ee
where the quantized electric flux $\hat f'$ and magnetic flux $\hat g'$ are given \cite{Lu:2000ys}, respectively, as
\be\label{quantizedem}
\hat f' = \frac{p'}{\sqrt{p'^{2} + \frac{q'^{2} + n'^{2}}{g^{2}_{s}}}}, \qquad \hat g' = \frac{q'}{n'}.
\ee
The above three integers $n', p', q'$ without a common divisor count the multiplicity of D3 branes, the quantized electric flux and the quantized magnetic flux, respectively. In many aspects, the discussion of the present case follows almost the same as that given in the trick used in the previous section when the magnetic flux $\hat g'$ there is taken to be finite rather than infinity.  Even so, there is still an important difference in that there the $\hat f'$ and $\hat g'$ are independent of each other while in the present case this is not so as indicated in (\ref{quantizedem}).  This will have effects on the actual pair production rate as will be seen later on.

The closed string cylinder interaction amplitude between the two D3 branes placed parallel at a separation $y$ and carrying their respective worldvolume flux in the form as given in (\ref{flux}) and (\ref{flux'}) has been computed by one of the present authors along with his collaborator(s) in \cite{Lu:2009au, Lu:2017tnm, Lu:2018suj, Jia:2019hbr} and is given as
\bea \label{cylinder-amplit-new}
\Gamma_{3, 3} &=& \frac{4 i  n' V_4 |\hat f - \hat f'| |\hat g - \hat g'|}{(8\pi^2 \alpha')^2} \int_0^\infty \frac{d t}{t^3} e^{- \frac{y^2}{2\pi \alpha' t}} \frac{\theta_1^2\left(\left.\frac{i \bar\nu_0 - \nu_1}{2}\right| it \right) \theta_1^2 \left(\left.\frac{i\bar\nu_0 + \nu_1}{2}\right| it\right)}{\eta^6 (it) \theta_1 (i\bar\nu_0| it) \theta_1 (\nu_{1} | it)}\nn
&=&\frac{4  n' V_4 (\cosh \pi\bar\nu_0 - \cos\pi\nu_1)^2 \sqrt{ (1 - \hat f^2) (1 - \hat f'^{2}) (1 + \hat g^2)(1 + \hat g'^{2})}}{(8\pi^2 \alpha')^2} \int_0^\infty \frac{d t}{t^3} e^{- \frac{y^2}{2\pi \alpha' t}} \nn
 &\times& \prod_{n= 1}^{\infty}  \frac{[1 - 2 e^{-\pi \bar\nu_0}  |z|^{2n} \cos\pi\nu_{1} + e^{- 2 \pi \bar\nu_0} |z|^{4n}]^2 [1 - 2 e^{\pi \bar\nu_0}  |z|^{2n} \cos\pi\nu_1 + e^{ 2 \pi \bar\nu_0} |z|^{4n}]^2}{(1 - |z|^{4n})^4 (1 - 2  |z|^{2n}\cosh2\pi\bar\nu_0 + |z|^{4n})(1 - 2  |z|^{2n} \cos\pi \nu_1 + |z|^{4n})},\nn
\eea
where $|z| = e^{- \pi t} < 1$ and the parameters $\bar\nu_{0} \in [0, \infty)$ and $\nu_{1} \in[0, 1)$ are determined, respectively, via
\be\label{nu-parameter-new}
\tanh \pi \bar\nu_{0} = \frac{|\hat f - \hat f'|}{1 - \hat f \hat f'}, \quad \tan\pi \nu_{1} = \frac{|\hat g - \hat g'|}{1 + \hat g \hat g'}.
\ee
By the same  token as we did in the previous section, the present amplitude is also positive for reasonably large brane separation $y$ and therefore the interaction is also attractive.  As such, the underlying system does not preserve any supersymmetry. For small $y$ for which the small $t$ integration in the amplitude becomes important, each factor in the infinite product in the integrand can be negative in a similar spirit as discussed in the previous section and so the sign of the infinite product becomes indefinite again, signaling new physics to occur.  This becomes manifest in terms of the open string variable.  The corresponding open string one-loop  annulus amplitude can be obtained once again by a simple Jacobi-transformation via $t \to 1/t$ to the right side of the first equality in (\ref{cylinder-amplit-new})  in a similar fashion as we did in the previous section  and the resulting amplitude is
\bea \label{annulus-amplit-new}
\Gamma_{3,3} &=&   \frac{4  n'   V_4 |\hat f - \hat f'| |\hat g - \hat g'|}{(8 \pi^2 \alpha')^2} \int_0^\infty \frac{d t}{t}\frac{(\cosh \pi \nu_1 t - \cos\pi \bar\nu_0 t)^2}{\sin\pi \bar\nu_0 t \,\sinh\pi \nu_1 t}\,
e^{- \frac{y^2 t}{2\pi \alpha'}}  \nn
&\,&\times \prod_{n = 1}^{\infty} \frac{|1 - 2 |z|^{2n} e^{- i \pi \bar\nu_{0} t} \cosh \pi \nu_{1} t  + |z|^{4n} e^{- 2 i \pi \bar\nu_{0} t}|^{4}}{(1 - |z|^{2n})^{4} [1 - 2 |z|^{2n} \cosh 2\pi \nu_{1} t + |z|^{4n}][1 - 2 |z|^{2n} \cos2\pi\bar\nu_{0} t + |z|^{4n}]},
\eea
where again $|z| = e^{- \pi t} < 1$. Here too, except for the factor $\sin \pi \bar\nu_{0} t$ in the denominator of the above integrand, all other factors are positive.  For large $t$, the integrand blows up
when $y < \pi \sqrt{2 \nu_{1} \alpha'}$ since it behaves
\be\label{tachyon-new}
e^{- \frac{y^2 t}{2\pi \alpha'}} e^{\pi  \nu_{1} t} = e^{- 2\pi \alpha' t \left[\frac{y^{2}}{(2\pi \alpha')^{2}} - \frac{\nu_{1}}{2 \alpha'}\right]},
\ee
where the factor $\nu_{1}/2$ in the second term of the square bracket in the exponential is the tachyonic shift.   The quantity $y^{2}/(2\pi \alpha')^{2} - \nu_{1}/(2 \alpha')$ as before defines the lowest energy square or the effective mass square of the open string, connecting our D3 and the other D3, which we will discuss later on.  The aforementioned factor $\sin\pi \bar \nu_{0} t$, as before, gives rise to an infinite number of simple poles of the integrand along the positive t-axis at $t_{k} = k /\bar\nu_{0}$ with $k = 1, 2, \cdots$, once again signaling the decay of the underlying system via the so-called open string string pair production.  The decay rate can be obtained following the procedure given in the previous section as
\be\label{decay-rate-new}
{\cal W} = \frac{8 n' |\hat f - \hat f'| |\hat g - \hat g'|}{(8 \pi^{2} \alpha')^{2}} \sum_{k = 1}^{\infty} \frac{(- )^{k - 1} }{k} \frac{\left[\cosh \frac{ k \pi \nu_{1}}{\bar\nu_{0}} - ( - )^{k}\right]^{2}}{\sinh
\frac{ k \pi \nu_{1}}{\bar \nu_{0}}} \, e^{- \frac{k y^{2}}{2\pi \bar\nu_{0} \alpha'}} Z_{k} (\bar\nu_{0}, \nu_{1}),
\ee
where $Z_{k} (\bar\nu_{0}, \nu_{1})$ is again given by (\ref{Zk}) but with the present $\nu_{1}$.  The open string pair production rate is just the $k = 1$ leading term of the above rate, following \cite{nikishov}, as
\be\label{pp-rate-new}
{\cal W}^{(1)} = \frac{8 n' |\hat f - \hat f'| |\hat g - \hat g'|}{(8 \pi^{2} \alpha')^{2}}\, \frac{\left[\cosh \frac{  \pi \nu_{1}}{\bar\nu_{0}} + 1\right]^{2}}{\sinh
\frac{ \pi \nu_{1}}{\bar \nu_{0}}} \, e^{- \frac{ y^{2}}{2\pi \bar\nu_{0} \alpha'}} Z_{1} (\bar\nu_{0}, \nu_{1}).
\ee
This rate is the starting point of discussion in the present section.  Previously, the rate was discussed for $\hat f' = 0, \hat g' = 0$, for example, in \cite{Lu:2009au, Lu:2017tnm, Lu:2018nsc}.  Here we will focus on the quantized electric flux $\hat f'$ and the quantized magnetic flux $\hat g'$ as given in (\ref{quantizedem}), respectively.   There are actually three sub-cases to consider: 1) $p' = 0, q' \neq 0$ ($\hat f' = 0, \hat g' \neq 0$); 2) $p' \neq 0, q'  = 0$ ($\hat f' \neq 0, \hat g' = 0$); 3) $p' \neq 0, q' \neq 0$ ($\hat f' \neq 0, \hat g' \neq 0$). Note that once again both the earthbound  electric flux $\hat f = 2\pi \alpha' e E \sim 10^{- 21} \ll 1$ and the earthbound magnetic flux $\hat g = 2 \pi \alpha' e B \sim 10^{- 21} \ll 1$ as discussed in the previous section.  We now discuss each of the sub-cases in order.

\medskip
\noindent
{\bf Sub-case 1)} $p' = 0, q' \neq 0$: For this sub-case, we have from (\ref{quantizedem})
\be\label{subc1}
\hat f' = 0,  \quad \hat g' = \frac{q'}{n'}.
\ee
So for any finite $q'$, we have $\hat g'$ being finite due to $n' \ge 1$.  Since $\hat f = 2 \pi \alpha' e E \ll 1$ and $\hat g = 2 \pi \alpha' e B \ll 1$, we have from (\ref{nu-parameter-new})
\be\label{subc1-np}
\bar \nu_{0} \approx 2 \alpha' e E \sim 10^{- 22} \ll 1, \quad \tan \pi \nu_{1} = \frac{q'}{n'},
\ee
where we assume $q' > 0$.  We therefore have $Z_{1} (\bar\nu_{0}, \nu_{1}) \approx 1$.  Given this and  $\nu_{1}/\bar \nu_{0} \gg 1$ due to $\nu_{1}$ being finite, the rate (\ref{pp-rate-new}) becomes in the present sub-case as
\be\label{subc1-pprate}
{\cal W}^{(1)} \approx \frac{q' e E}{8 \pi^{3} \alpha'} e^{- \frac{\pi m^{2}_{\rm eff}}{e E} },
\ee
where the effective mass is again given by (\ref{effm}) with the present $\nu_{1}$. Just like the case of the D3/D1, the contribution to the present rate comes also from the pair of the massive charged/anti charged vector polarizations with their lowest effective mass following the same discussion as given in the previous section due to the finite $\nu_{1}$ while the other 15 pairs of massive charged/anti charged polarizations have each an almost vanishing contribution in comparison with this pair. This rate looks almost the same as that for the D3/D1 system reported in \cite{Lu:2019ynq} except for one minor difference. For this,  we have there $\nu_{1} \to 1/2$ while for the present case we have a finite $\nu_{1}$ in the range of $(0, 1/2)$ for any finite $q'$ along with $n' \ge 1$.  The larger $q'$ and $\nu_{1}$ are, the larger the rate is. The most efficient is to take $n' = 1$ and this will give the largest $\nu_{1}$ for given $q'$.  Just like the D3/D1 system discussed earlier, the pair production for this sub-case is also insignificant before the tachyon condensation due to the applied electric flux being too small.  \\
\medskip
\noindent
{\bf Sub-case 2)} $p' \neq 0, q' = 0$: For this sub-case, we have from (\ref{quantizedem})
\be\label{subc2}
\hat f' = \frac{g_{s} p'}{\sqrt{(g_{s} p')^{2}  + n'^{2}}}, \quad \hat g' = 0.
\ee
For not too large $p'$, say $p' = 10$, the smaller $n'$ is, the larger $\hat f'$. So we take $n' = 1$ for this sub-case. We then have $\hat f' \approx g_{s} p'  = 0.1\gg \hat f \sim 10^{- 21} $ if we take $g_{s} = 10^{-2}$ as before.   With this and $\hat g' = 0$, we have from (\ref{nu-parameter-new})
\be\label{subc2-np}
\tanh \pi \bar\nu_{0} = \hat f' = g_{s} p' = 0.1, \quad \nu_{1} = \frac{\hat g}{\pi} = 2\alpha' e B \sim 10^{- 22} \ll 1.
\ee
So we  have $\bar\nu_{0} \approx g_{s} p'/\pi = 1/(10 \pi) \ll 1$ and $\pi \nu_{1}/\bar \nu_{0} \sim 10^{-20} \ll 1$.  The former implies $Z_{1} (\bar\nu_{0}, \nu_{1}) \approx 1$, indicating only the lowest mass  modes of the open string and the anti open string connecting our D3 and the other D3 to contribute possibly to the pair production.  These are the eight bosons ($8_{\rm B}$) and eight fermions ($8_{\rm F}$) for either the open string or the anti open string, giving a total of 16 pairs of charged/anti-charged modes.  To the brane observer,  the above $8_{\rm B}$ and $8_{\rm F}$ modes of the open string correspond to one of the two broken generators of the original 4-dimensional N = 4 U(2) super Yang-Mills  when $y = 0$  as $U(2) \to U(1) \times U(1)$ when $y \neq 0$.  The $8_{\rm B}$ and $8_{\rm F}$ modes of the anti open string correspond to the other broken generator and each carries the opposite unity charge with respective to those modes of the open string under the unbroken $U(1)$ associated with our own D3.  Note that the underlying supersymmetries are also broken by the presence of fluxes, in particular the quantized flux $\hat f'$.  Further with $\nu_{1}/\bar\nu_{0} \sim 10^{- 21}$,  the pair production rate (\ref{pp-rate-new}) for the present sub-case becomes
\be\label{subc2-pprate}
{\cal W}^{(1)} \approx \frac{2 (e E')^{2}}{\pi^{3}} e^{- \frac{\pi m^{2}}{e E'}},
\ee
where we have used $\hat f' = 2 \pi \alpha' e E'$ with $e E' = g_{s} p' / (2\pi \alpha') = M^{2}_{s} /(20 \pi)$ and $m = y/(2\pi \alpha')$ the lowest mass for the above $8_{\rm B}$ and $8_{\rm F}$ modes of the open string or the anti open string.  In the presence of the applied  magnetic flux $\hat g$, we expect the mass splittings for these charged modes \cite{Ferrara:1993sq} but these splittings, unlike the previous sub-case and the case discussed in the previous section,  are insignificant, as indicated in the above rate, because of $\nu_{1}/\bar\nu_{0} \ll 1$.  So we have here all 16 pairs of charged/anti-charged massive modes contributing to the pair production rate (\ref{subc2-pprate}).

Note also that the rate itself is independent of the applied electric and magnetic fields, similar to the case discussed in the previous section.  The present rate is new and the actual pair production, by the same token as discussed in the previous section, can be estimated from the requirement  $e E' \ge m^{2}$.   Once again, given  $e E'  = M^{2}_{s}/(20 \pi)$ being not much smaller than the corresponding string scale, the above condition can be satisfied easily.  With $m = y/(2 \pi \alpha') = M^{2}_{s} y /(2 \pi)$, this gives the condition for the brane separation as $ y \le  y_{0} = l_{s} \sqrt{\pi/5} \approx 0.79\, l_{s} $.  In other words, when the D3 carrying the quantized electric flux $\hat f' = g_{s} p' = 2\pi \alpha' e E'$ comes close to our D3 at the brane separation of $y \le y_{0}  \approx 0.79 \,l_{s}$,  the open string pair production rate is large enough to give sizable pair production.  For the present sub-case, it does not appear that we have a tachyon condensation issue.  However, due to $y_{0} = 0.79\, l_{s}$ being substring scale, we need to worry about the other condition with which our D3 and the non-threshold bound state (D3, (F, D1)) can both be taken as rigid to validate our computations as mentioned in the previous section and discussed in \cite{Jia:2019hbr}.  This gives the requirement for the probe distance $y \gg (3 \pi g_{s})^{1/4} l_{s} \approx 0.55\, l_{s} $.  Since the $y_{0} \approx 0.79\, l_{s}$ is not much larger than $0.55\, l_{s}$,  there may be issues associated with the validity of the rate computations for this sub-case though it may be a preferred choice  due to all 16 pairs contributing to the pair production\footnote{It may be indeed so to the leading order approximation but the rate computations presented here may also have  modifications and could be invalid since the branes under consideration can no longer be taken as rigid  and the effects of curved spacetime due to the presence of branes at the probe distance may need to be taken into consideration.}.

\medskip
\noindent
{\bf Sub-case 3)} $p' \neq 0, q' \neq 0$: This appears to be the most interesting sub-case. We have, from (\ref{quantizedem}), both $\hat f' \neq 0$ and $\hat g' \neq 0$.  We again take not too large $p'$, say, $p' = 10$.  With $g_{s} = 10^{-2}$, noting $q' \ge 1, n' \ge 1$, we have from (\ref{quantizedem})
\be\label{subc3}
\hat f' \approx \frac{g_{s} p'}{\sqrt{n'^{2} + q'^{2}}} < 0.1.
\ee
We have then from (\ref{nu-parameter-new})
\be\label{subc3-np}
\bar\nu_{0} \approx \frac{\hat f'}{\pi} < \frac{1}{10 \pi}, \quad \tan \pi \nu_{1} \approx \hat g' = \frac{q'}{n'},
\ee
noting $\hat f \sim \hat g \sim 10^{-21} \ll 1$.  Since $\bar \nu_{0} < 1/(10 \pi)$ and $|z_{1}| = e^{- \pi/\bar\nu_{0}} < e^{- 10 \pi^{2}} \ll 1$, we have $Z_{1} (\bar \nu_{0}, \nu_{1}) \approx 1$, once again indicating that at most the $8_{\rm B}$ and $8_{\rm F}$ modes from the open string and  the same modes from the anti open string contribute to the pair production. From (\ref{subc3-np}),  $\nu_{1}$ is less than $1/2$ but they are on the same order in general.  Combining all these,  we have the present pair production rate from (\ref{pp-rate-new}) as
\be\label{subc3-pprate}
{\cal W}^{(1)} \approx \frac{ q'  e E' }{8 \pi^{3} \alpha' } \, e^{- \frac{\pi m^{2}_{\rm eff}}{e E'}},
\ee
where $m_{\rm eff}$ is given by eq.(\ref{effm}) with the present $\nu_{1}$ and we have also used $\hat f' = 2\pi \alpha' e E'$ in the above. This rate, as in sub-cases 1) and 2), is now the number of pairs produced per unit time per unit D3 brane volume.  In obtaining the above rate from (\ref{pp-rate-new}), we have considered $\pi \nu_{1}/\bar\nu_{0} > 10 \pi^{2} \nu_{1} \sim 5 \pi^{2} \gg 1$  such that we can replace both $\cosh \pi \nu_{1}/\bar\nu_{0}$ and $\sinh\pi\nu_{1}/\bar \nu_{0}$ by the exponential factor $e^{\pi\nu_{1}/\bar\nu_{0}}$ in the rate formula. This remains true indeed given that $\nu_{1}$ is in general on the order of $1/2$.  For example, for the sample case considered in the following, we have $\pi \nu_{1}/\bar \nu_{0} = 5 \pi^{2}/\sqrt{2} \approx 35 \gg 1$.   So, just like the sub-case 1) and the case discussed in the previous section,   only the pair of the charged/anti-charged vector polarizations with the above lowest effective mass $m_{\rm eff}$ contributes to the above rate while the other 15 pairs of charged/anti-charged polarizations have their comparatively vanishing contributions to this rate.

For a large rate, we want a possible large $\hat f'$ or $e E'$ while also keeping $\nu_{1}$ as large as possible.  A sample choice of both from (\ref{subc3}) and (\ref{subc3-np}) is to take $q' = n' = 1$ and this gives $\hat f' = g_{s} p' /\sqrt{2} = 1/(10 \sqrt{2})$ and $\nu_{1} = 1/4$.  So we have $e E' = \hat f'/(2\pi \alpha') = M^{2}_{s} /(20 \pi \sqrt{2})$.  To actually produce the pair, as before we need to have $e E' \ge m^{2}_{\rm eff}$ which gives the brane separation, from (\ref{effm}) and with $\nu_{1} = 1/4$, as $y \le y_{0} + 0.1 \, l_{s}$ with $y_{0} = (\pi/\sqrt{2}) \, l_{s} \approx 2.2 \, l_{s}$ the brane separation at which the tachyon condensation starts to work.  In other words, we expect to produce  the pairs actually once the brane separation falls in the range of $y_{0} < y \le y_{0} + 0. 1\, l_{s}$ before the tachyon condensation occurs.

Note that the present rate (\ref{subc3-pprate}) looks essentially the same as that (\ref{eff-pprate}) given in the previous section for the system of D3/(F, D1).  Both of the cases have the same range of $\Delta y = 0.1\, l_{s}$ ($y_{0} < y \le y_{0} + 0.1\, l_{s}$) for which both the rates can actually give rise to the pair production.  However, there is a small
difference in that the present $y_{0} \approx 2.2 \, l_{s}$ while in the previous section $y_{0}  = \pi\, l_{s} = 3.1 \, l_{s}$.  In this aspect, the approximation of taking the branes as rigid in our rate computations serves a bit better for the D3/(F, D1) system considered in the previous section than the present D3/(D3, (F, D1)) one.

\section{Discussion and Conclusion}
In this paper,  we have considered two new systems, namely,  D3/(F, D1) and D3/(D3, (F, D1)), for the purpose of seeking more and better possibilities to actually give rise to the open string pair production before the corresponding tachyon condensation occurs over the previously studied system of D3/D1 reported in \cite{Lu:2019ynq}.  In this aspect,  we indeed achieve the goal for either of these two new systems in this work.

For the D3/(F, D1) system,  the quantized electric flux gives a much larger electric field $e E' \sim M^{2}_{s}/(20 \pi)$ with $M_{s}$ the string scale and being larger than a few TeV, see  \cite{Berenstein:2014wva}, for example.  While for the D3/D1 system,  the corresponding electric field $e E$ is  a much smaller earthbound laboratory one whose present largest value is only on the order of $10^{- 8} \,m^{2}_{e}$ with $m_{e} = 5.1 \times 10^{- 7}$ TeV, the electron mass.  As such, the former gives us a much larger chance of giving rise to the open string pair production which has the contribution, in both cases, only from the pair of charged/anti-charged vector polarizations with their effective mass  given by (\ref{effm}) with $\nu_{1} = 1/2$ and $m = y/(2\pi \alpha')$.  To actually produce the pairs, we need to have $e E' \ge m^{2}_{\rm eff}$ for the former and $e E \ge m^{2}_{\rm eff}$ for the latter.  The former gives the brane separation in the range of $y_{0} < y \le y_{0} + 0.1\, l_{s}$, independent of the string scale,  while the latter gives $y_{0} < y \le y_{0} + 2 \pi (m_{e}/M_{s})^{2}\, l_{s}$ whose range depends on the string scale, with $y_{0} = \pi \, l_{s}$ the brane separation at which tachyon condensation occurs.   For example,  if we take $M_{s} \sim$  a few TeV, i.e. its possible lower bound,  the range for latter case is $y_{0} < y \le y_{0} + 10^{-21}\, l_{s}$, much less than the former one.  This makes the production of the pairs almost impossible for latter case before the tachyon condensation occurs.

For the D3/(D3, (F, D1)) system, we have considered three sub-cases in section 3.  The sub-case 1) looks almost identical to that of the D3/D1 system while the sub-case 3) looks almost the same as that for the D3/(F, D1) system.  For each of the case considered in \cite{Lu:2019ynq} and the case/sub-cases in this paper,  except for the sub-case 2) considered in the previous section, the corresponding pair production rate has the only contribution from the pair of the charged/anti-charged vector polarizations with the lowest mass $m_{\rm eff}$ given in (\ref{effm}) with $\nu_{1}$ the same order of $1/2$ in general. However, for the sub-case 2), the story is different.  All 16 pairs of charged/anti-charged polarizations, including 5 pairs of charged/anti-charged massive scalars, 4 pairs of charged/anti-charged massive spinors (each spinor has two polarizations) and one pair of charged/anti-charged massive vectors (each massive vector has three polarizations),  contribute to the pair production rate, with each of these modes or polarizations having basically the same mass $m = y/(2\pi \alpha')$.  However, for this sub-case, the brane separation at which the pair production occurs most likely is $y \le 0.79\, l_{s}$ which is not much larger than the required $y = 0.55\, l_{s}$ to validate our rate computations. So this sub-case may be interesting and useful but the present discussion is not conclusive.

In conclusion,  we can indeed have a large pair production for either of D3/(F, D1) or the sub-case 3) of D3/(D3, (F, D1)).  As discussed in the previous section, even so the D3/(F, D1) seems in a bit better position than the D3/(D3, (F, D1)) in the sense that the former has a bit larger $y_{0} = \pi\l_{s}$ (the latter has $y_{0} = (\pi/\sqrt{2}) \, l_{s}$), at which the corresponding tachyon condensation occurs, to validate the rate computations.

Given the large pair production, the natural questions is: Can either of the above be detectable if we are lucky to have this process occurring indeed? Unless there is a special technology advance for detecting this process, the answer is most likely no since this large rate occurs in a too short duration of the order of one tenth of string scale before the underlying tachyon condensation occurs. This almost transient process is due to the small brane separation range of $0.1\, l_{s}$ and the brane interaction being attractive.

In spite of the above, as mentioned in the Introduction, this transient process, if it occurs  indeed, can give rise to the high energy radiation in a very short period of time and this may provide an alternative mechanism for the reheating process if it occurs at the early Universe or it may be used to provide explanation to the observed $\gamma$-ray burst if
occurs at the later time of our universe.  Exploring either of these possibilities may be interesting and worthwhile and we hope in the near future to come back to address them.


\section*{Acknowledgments}
We acknowledge the support by grants from the NNSF of China with Grant No: 11235010 and 11947301.

\end{document}